\documentclass[a4paper,onecolumn,11pt, unpublished]{quantumarticle}

\pdfoutput=1

\usepackage[usenames,dvipsnames]{xcolor}
\usepackage{physics}
\usepackage{bbold}
\usepackage{comment}
\usepackage{dsfont}
\usepackage{booktabs}

\usepackage[utf8]{inputenc}

\DeclareFixedFont{\ttb}{T1}{txtt}{bx}{n}{10} 
\DeclareFixedFont{\ttm}{T1}{txtt}{m}{n}{10}  

\definecolor{deepblue}{rgb}{0,0,0.5}
\definecolor{deepred}{rgb}{0.6,0,0}
\definecolor{deepgreen}{rgb}{0,0.5,0}

\usepackage{listings}

\newcommand\pythonstyle{\lstset{
language=Python,
basicstyle=\ttm,
morekeywords={self},              
keywordstyle=\ttb\color{deepblue},
emph={MyClass,__init__},          
emphstyle=\ttb\color{deepred},    
stringstyle=\color{deepgreen},
frame=tb,                         
showstringspaces=false
}}

\lstnewenvironment{python}[1][]
{
\pythonstyle
\lstset{#1}
}
{}

\newcommand\pythoninline[1]{{\pythonstyle\lstinline!#1!}}

\providecommand{\ket}[1]{\lvert #1 \rangle}

\newcommand{\dager}{^{\dagger}}
\renewcommand{\a}{\hat{a}}

\newcommand{\x}{\hat{x}}

\newcommand{\p}{\hat{p}}

\renewcommand{\H}{\hat{H}}

\newcommand{\half}{\frac{1}{2}}

\newcommand{\eye}{\mathbb{1}}

\usepackage{lipsum}

\usepackage{graphicx}
\usepackage{dcolumn}
\usepackage{bm}
\usepackage[parse-numbers=false]{siunitx}
\usepackage{upgreek}

\begin{document}

\title{QuGIT: a numerical toolbox for Gaussian quantum states} 


\author{I. Brand\~{a}o}
\email{igorbrandao@aluno.puc-rio.br}
\affiliation{Departamento de F\'{i}sica, Pontif\'{i}cia Universidade Cat\'{o}lica do Rio de Janeiro,  22451-900 Rio de Janeiro, RJ, Brazil}
\orcid{0000-0003-4710-7721}

\author{D. Tandeitnik}
\email{tandeitnik@aluno.puc-rio.br}
\affiliation{Departamento de F\'{i}sica, Pontif\'{i}cia Universidade Cat\'{o}lica do Rio de Janeiro,  22451-900 Rio de Janeiro, RJ, Brazil}
\orcid{0000-0003-3276-9335}

\author{T. Guerreiro}
\email{barbosa@puc-rio.br}
\affiliation{Departamento de F\'{i}sica, Pontif\'{i}cia Universidade Cat\'{o}lica do Rio de Janeiro,  22451-900 Rio de Janeiro, RJ, Brazil}
\orcid{0000-0001-5055-8481}


\begin{abstract}
Simulating quantum states on a classical computer is hard, typically requiring prohibitive resources in terms of memory and computational power. Efficient simulation, however, can be achieved for certain classes of quantum states, in particular the so-called Gaussian quantum states of continuous variable systems. In this work we introduce QuGIT - a python numerical toolbox based on symplectic methods specialized in efficiently simulating multimode Gaussian states and operations. QuGIT is exact, requiring no truncation of Hilbert space, and provides a wide range of Gaussian operations on arbitrary Gaussian states, including unitaries, partial traces, tensor products, general-dyne measurements, conditional and unconditional dynamics. To illustrate the toolbox, several examples of usage relevant to quantum optics and optomechanics are described.
\end{abstract} 

\maketitle

\section{Introduction}\label{sec:introduction}



Simulating arbitrary quantum systems is a difficult task for classical computers. The memory necessary to store quantum states in cache and the complex computations required to emulate their dynamics generally scales exponentially with the number of modes for finite-dimensional Hilbert spaces \cite{qutip1}. 
For infinite-dimensional systems the situation is even worse since dynamics cannot be reproduced with limited memory in general, regardless of the number of modes.
To overcome these limitations, one often works with truncated Hilbert space dimensions, which works well if the system has a small number of modes and excitations. Dimensional truncation is commonly used in a variety of quantum numerical packages \cite{julia_toolbox, Tan1999}, and most prominently in the Quantum Toolbox in Python (QuTiP) \cite{qutip1, qutip2}, which has been extensively used to simulate a wide range of systems from parametric amplifiers and frequency converters to superconducting qubits and quantum Monte-Carlo trajectories in cavity QED  \cite{qutip1}.


While dimension truncation offers a viable path to simulating a myriad of quantum systems, it is desirable to have complementary tools which are exact. Of particular interest to us is the special class of continuous variable Gaussian quantum states, which comprise a number of interesting situations, particularly in quantum optomechanics experiments involving strong coherent states \cite{Aspelmeyer2014} and nanomechanical resonators interacting with thermal environments possessing large occupation numbers \cite{Thesis_Delic2019, DeLosRios2020}, both intractable using truncated Hilbert spaces. Quantum optical circuits involving Gaussian states with a large number of modes such as in Boson sampling \cite{Zhong2020} and photonic quantum computing \cite{bartolucci2021} are also examples for which exact methods are desirable. For these needs we have developed the \textbf{Qu}antum \textbf{G}aussian \textbf{I}nformation \textbf{T}oolbox - or QuGIT for short - a python numerical toolbox for Gaussian quantum information applications.

Gaussian states are completely characterized by the first and second moments of their phase-space distribution \cite{Lloyd2012}. This lifts the requirement of dealing with the infinite-dimensional objects characteristic of continuous variable systems, requiring only a finite covariance matrix for the complete and exact description of the system's state and dynamics. Consequently, memory requirements grow only quadratically with the number of modes $ N $; specifically only $4N^2+2N$ double-precision floating point numbers in memory are needed to represent an $N$-mode Gaussian state. 
QuGIT builds upon the formalism of Gaussian quantum states to implement a series of useful tools for continuous variable quantum systems. 

This paper is organized as follows. In Section \ref{sec:theory}, we present a brief introduction to Gaussian quantum states, its symplectic representation and Gaussian-preserving dynamics. The framework of QuGIT and how the toolbox is divided is presented in Section \ref{sec:framework}, while a detailed discussion is provided in Section \ref{sec:examples} through a series of illustrative examples covering a wide range of the toolbox capabilities. The associated codes for these simulations are presented in Appendix \ref{appendix:example_codes}, and the list of QuGIT's built-in functions is given in Appendix \ref{sec:gaussian_states_methods}. In Section \ref{sec:performance} we discuss the performance of QuGIT to simulate dynamics in comparison to QuTiP. Section \ref{sec:conclusions} concludes with final considerations on the current state of the toolbox and how it contributes to the list of already existing packages for simulating quantum systems.

\section{Brief review of Gaussian states and notation} \label{sec:theory}

\subsection{Gaussian states} \label{sec:gaussian_state_theory}
QuGIT is based on the formalism of \textit{continuous variable quantum systems} whose states live in an infinite-dimensional Hilbert space described by observables with continuous spectra. In the remaining of this text we follow Refs. \cite{Lloyd2012, alessioserafini2017, Grimmer2018, Genoni2016} and a brief summary of the main properties regarding continuous variable (CV) systems is provided in this section. 

An $N$-mode CV state is characterized by annihilation and creation operators $\a_j$ and $\a_j\dager$ ($j = 1 \ldots N$) obeying the standard bosonic commutation relations  $\comm{\a_j}{\a_k\dager} =  \eye \, \delta_{j,k}$.
We choose to work in units where $\hbar=2$, such that the canonical observables associated to each mode are defined as $\x_j \equiv \a_j\dager + \a_j$ and $\p_j \equiv i\big(\a_j\dager - \a_j\big)$. These can be interpreted as the position and momentum operators of a harmonic oscillator or field quadratures of an optical mode.
In the remaining of this work we shall refer to $ \x_j, \p_j $ as \textit{quadrature operators} and it follows from the bosonic commutation relations that $\comm{\x_j}{\p_k} = 2i\eye \, \delta_{j,k}$ and $\comm{\x_j}{\x_k} = \comm{\p_j}{\p_k} = 0$.

A convenient representation of quadrature operators is via a vector of the form $\bm{\hat{r}} = \big( \x_1, \p_1, \x_2, \p_2, \ldots \big)^T$. This allows one to write the canonical commutation relations in compact notation, 
\begin{align}
    \comm{\hat{r}_j}{\hat{r}_k} = 2i \, \Omega_{j,k} \  , \ \ \  \Omega \equiv \bigoplus_{k=1}^N \begin{bmatrix}
    0 & 1 \\
    -1 & 0
    \end{bmatrix} \, . \label{eq:commutation_relations_quadratures}
\end{align}
where $\Omega$ is the so-called symplectic form matrix.

The continuous spectra of the quadrature operators span a real symplectic space \cite{Lloyd2012} referred to as \textit{phase-space}. This allows for a convenient representation of arbitrary continuous variable states: 
the complete information contained in an $N$-mode bosonic quantum state $\rho$ can be represented as a quasi-probability distribution in a $2N$-dimensional phase-space. In this work we are interested in the special class of continuous variable states for which the phase-space distribution assumes a Gaussian form. 
These so-called \textit{Gaussian states} can be completely described by the first and second moments of their quadrature operators. 

Let $\rho_G$ be an $N$-mode Gaussian state; its first moment is given by the quadrature vector $\bm{R} \in \mathbb{R}^{2N}$, defined as
\begin{equation}
    \bm{R} \equiv \langle \bm{\hat{r}} \rangle = \tr(\rho_{G} \bm{\hat{r}}) \, ,
\end{equation}
while the second moments can be arranged into a real symmetric \textit{covariance matrix} $V \in \mathbb{R}^{2n\times2n}$, with entries given by
\begin{equation}
    V_{j,k} = \frac{1}{2}\langle \hat{r}_j \hat{r}_k + \hat{r}_k \hat{r}_j\rangle - \langle \hat{r}_j \rangle \langle \hat{r}_k \rangle \, .
\end{equation}
These are the basic data elements of QuGIT.

\subsection{Gaussian preserving dynamics} \label{sec:gausian_dynamics_theory}

Physical transformations that map Gaussian states to Gaussian states are called \textit{Gaussian-preserving}. To be a \textit{Gaussian-preserving unitary} the operator $\hat{S} = \exp(-i\H_0/2)$ must be generated by a Hamiltonian $\H_0$ that is \textit{at most} quadratic in the quadrature operators and thus its most general form is
\begin{equation}
    \H_0 = \frac{1}{2}\bm{\hat{r}}^T H(t) \bm{\hat{r}} + \bm{\alpha}_H^T \bm{\hat{r}} \label{eq:gaussian_Hamiltonian}
\end{equation}
\noindent where $H(t)$ is a time-dependent real symmetric $2N\times2N$ matrix and $\bm{\alpha}_H$ is a real vector of length $2N$. Using the Heisenberg equations for the quadrature vector alongside its commutation relations given in Eq. \eqref{eq:commutation_relations_quadratures} we find that Gaussian unitaries can be described through an affine mapping of the quadrature operators
\begin{align}
    \dot{\bm{\hat{r}}} = \Omega H(t) \bm{\hat{r}} + \Omega\bm{\alpha}_H\eye \, .
\end{align}

The discussion on Gaussian-preserving transformations may also be generalized to account for open quantum system dynamics and continuous measurements.
The open dynamics is modelled by considering that the system of interest is weakly coupled to a large reservoir in an $M$-mode Gaussian state with first and second moments $\bm{R}_B=\bm{0}$ and $V_B$. 
The environment is assumed to satisfy the \textit{white noise} condition
\begin{equation}
    \langle \acomm{\bm{\hat{r}}_B(t)}{\bm{\hat{r}}_B^T(t')} \rangle = 2 V_B \ \delta(t-t')\, , 
\end{equation}
implying Markovian dynamics.
Moreover, the environment and system are allowed to interact through a quadratic coupling Hamiltonian, which preserves the Gaussian character of the global state, 
\begin{equation}
    \H_{\mathrm{int}} = \bm{\hat{r}}^T C \bm{\hat{r}}_B \, ,
\end{equation}
\noindent where $C$ is a $2N \times 2M$ real matrix describing the system-environment coupling and $\bm{\hat{r}}_B$ is the quadrature operator vector for the environment. 

Continuous measurements are modelled by a general-dyne detection scheme acting on the environment giving rise to a conditional dynamics on the system. Note this scheme preserves the Gaussianity of the system. 
The choice of measurement is characterized by the post-measurement covariance matrix of the monitored modes $V_M$, while the probability of measuring an outcome $\bm{r}_M$ from these modes follows a Gaussian probability density with mean $\bm{R}_B$ and covariance $(V_B+V_M)^{-1/2}$. For a detailed description of general-dyne measurements see Ref \cite{Genoni2016}.


After tracing out the environment, the conditional dynamics induced on the system by the total Hamiltonian $\H = \H_0 + \H_{\mathrm{int}}$ plus continuous general-dyne measurements is described by a set of stochastic Langevin equations together with a deterministic Riccati equation for the first moments,
\begin{align}
    \dot{V} &= AV + V^TA^T + D - \chi(V) \, , \label{eq:langevin}\\
    d\bm{R} &=  (A\bm{R} + \bm{N})dt + (V\mathcal{C}^T + \Gamma^T)d\bm{w} \, , \label{eq:lyapunov}
\end{align}
\noindent where $d\bm{w} = (V_B+V_M)^{-1/2}(\bm{r}_m - \bm{R}_B) $ is a Wiener process with $\langle \{d\bm{w},d\bm{w}^T\} \rangle = \eye dt$. The \textit{drift matrix} $A$, \textit{diffusion matrix} $D$ and \textit{driving term} $\bm{N}$ dictate the unconditional dynamics on the system whilst the monitoring of the system is introduced through the positive definite matrices $\mathcal{C}, \Gamma$ and $\chi(V)$,
\begin{equation}
  \begin{split}
    A      &= \Omega H + \half \Omega C \Omega C^T \, ,\\
    D      &= \Omega C V_B C^T \Omega^T \, ,\\
    \bm{N} &= \Omega\bm{\alpha}_H \, ,
  \end{split}
  \quad
  \begin{split}
    \chi(V)     &= (V\mathcal{C}^T + \Gamma^T) (\mathcal{C}V + \Gamma) \, ,\\
    \Gamma      &= (V_B+V_M)^{-1/2} V_B C \Omega \, , \\
    \mathcal{C} &= (V_B+V_M)^{-1/2} \Omega C\, .
  \end{split}
\end{equation}
We note that when $\mathcal{C} = \Gamma = 0$ the deterministic unconditional dynamics is recovered from the above equations and the Riccati equation reduces to a Lyapunov equation. 
We also observe that the effect of continuous monitoring introduces stochasticity only in the mean quadrature vector while the covariance matrix follows a deterministic dynamics.

The above matrices and vectors, alongside the initial state for the system are the basic data elements necessary for QuGIT to calculate arbitrary Gaussian quantum dynamics.




\section{QuGIT framework} \label{sec:framework}






\subsection{Emulating Gaussian states} \label{sec:gaussian_state_code}

The toolbox is able to emulate an arbitrary multi-mode Gaussian state, perform Gaussian operations and retrieve data from these simulations. This is achieved through the custom Python class \pythoninline{gaussian_state}. 
The attributes of this class encompass all the necessary information to characterize the states: their number of modes, mean quadrature vectors, covariance matrices and the associated symplectic form matrix, summarized in Table \ref{tab:properties_gaussian_state}. 

\begin{table}[!ht]
\centering
\caption{Attributes of the gaussian\_state class}
\label{tab:properties_gaussian_state}
\begin{tabular}{cc}
\hline
\hline 
Attribute & Description\\
    \midrule
R            & Mean quadrature vector \\
V            & Covariance matrix       \\
N\_modes     & Number of modes         \\
Omega        & Symplectic form matrix  \\
\hline\hline
\end{tabular}
\end{table}

Manipulations of the quantum state can be performed in two ways. One is through class methods that alter the class instance. The other is via homonym built-in functions that take as argument a \pythoninline{gaussian_state} and return a modified copy of the original class instance. 
Details regarding these operations are listed in the Appendix \ref{sec:gaussian_states_methods}.

We now present a first working example using the \pythoninline{gaussian_state} class. Consider a pair of two-mode Gaussian states. With the following code, we can find the their quantum fidelity: 
\begin{python}
import numpy as np
import quantum_gaussian_toolbox as qgt

R = np.array([1, 2, 3, 4])         # Mean quadrature vector for state 0
V = 10*np.eye(4)                   # Covariance matrix      for state 0
state_0 = qgt.gaussian_state(R, V) # Multimode Gaussian state

alpha = 1 - 2.0j                   # Coherent state 1 complex amplitude   
state_1 = qgt.coherent(alpha)      # Single mode Gaussian state

# Tensor product of two copies of a coherent state (state_1)
state_2 = qgt.tensor_product([state_1, state_1]) # Library function
state_1.tensor_product([state_1])                # Class method

F = qgt.fidelity(state_0, state_1) # Quantum fidelity between states
\end{python} 
The first lines of the code import the QuGit and Numpy packages. Note Numpy is necessary for the functioning of QuGIT. These imports and their aliases ``qgt.'' and ``np.'' are assumed in all examples from now on.
The creation of an arbitrary two-mode Gaussian state is achieved by initialising the first two moments of the state (declared as numpy.ndarrays) and passing these to the \pythoninline{gaussian_state} class constructor. 
We can initialise certain elementary Gaussian states - such as the vacuum, coherent, squeezed and thermal states - by a builtin function with its associated parameter. 
In the above example we can see the definition of a complex number (\pythoninline{alpha}) and the coherent state initialised with corresponding complex amplitude (\pythoninline{gqgt.coherent(alpha)}). 
Next, we calculate the tensor product of the initialised coherent state with itself. This tensor product can be achieved in two different ways: 
first with the QuGIT library function \pythoninline{qgt.tensor_product}, which takes a list of \pythoninline{gaussian_state} class instances and returns another instance with the tensor product. Second, we can make use of the \pythoninline{gaussian_state} class method to alter the variable \pythoninline{state_1} to the result of the tensor product. Finally, we calculate the quantum fidelity between \pythoninline{state_0} and \pythoninline{state_1}.

We note that while the above example contains some of the core mechanics of the class \pythoninline{gaussian_state}, there are numerous capabilities which can be achieved using the toolbox; see Appendix \ref{sec:gaussian_states_methods} for more details. 
These capabilities include Gaussian unitaries, phase-space representations, entanglement witnesses, general-dyne measurements and calculation of density matrices in different basis. 
By retrieving the density matrix on the number basis, the toolbox allows for almost effortless synergy with the widely used QuTiP package \cite{qutip1}.

\subsection{Simulating time evolution} \label{sec:gaussian_dynamics_code}
 
In addition to the state class, QuGIT contains the \pythoninline{gaussian_dynamics} class, for simulating Gaussian-preserving time evolution of a given initial state. 

The class attributes record the necessary information to calculate the time evolution, meaning the initial state (which is an instance of \pythoninline{gaussian_state}), the matrices appearing in equations \eqref{eq:langevin} and \eqref{eq:lyapunov} (defined as numpy.ndarrays) and the list of time-evolved states. 
The class methods are able to perform unconditional and conditional time evolution of multi-mode Gaussian states following the theory outlined in Section \ref{sec:gaussian_state_theory}, as well as calculate the semi-classical dynamics for the system and their steady states.

For the unconditional dynamics the class performs numerical integration of the deterministic differential equations and finds the associated steady state by solving the algebraic form of the Lyapunov and Langevin equations. 
Regarding the conditional case, the deterministic Riccati equation is solved using the same standard numerical integration, while a Monte Carlo method is employed to integrate the stochastic Langevin equation and yield quantum trajectories, induced by continuous monitoring, upon the Hilbert space of the system. 
Finally, for the semi-classical dynamics we generalize the unconditional case by considering the effect of stochastic forces acting on the Langevin equation with zero mean value and auto-correlation dictated by the diffusion matrix $D$; the resulting semi-classical dynamics gives rise to semi-classical trajectories.



\section{Examples} \label{sec:examples} 



We now proceed to discuss some illustrative examples using QuGIT. The code to each example is presented in the Appendix \ref{appendix:example_codes}. The associated lines of code used for plotting are omitted for simplicity, however, we note that throughout the remaining of this work, the matplotlib package \cite{matplotlib} is used for data visualization.

\subsection{Unitary field quadrature dynamics}

A standard plot in quantum optics textbooks \cite{Scully1997} is the unitary time development of the mean field quadrature and its corresponding variances for the electromagnetic field in various quantum states.
We begin with this simple example as the information required can be directly retrieved from the \pythoninline{gaussian_state} class attributes and serves to illustrate how one can use the toolbox to initialise states of interest, evaluate their unitary dynamics and visualize results via expectation values of canonical observables.

\begin{figure}[ht!]
    \centering
    \includegraphics[width=0.5\linewidth]{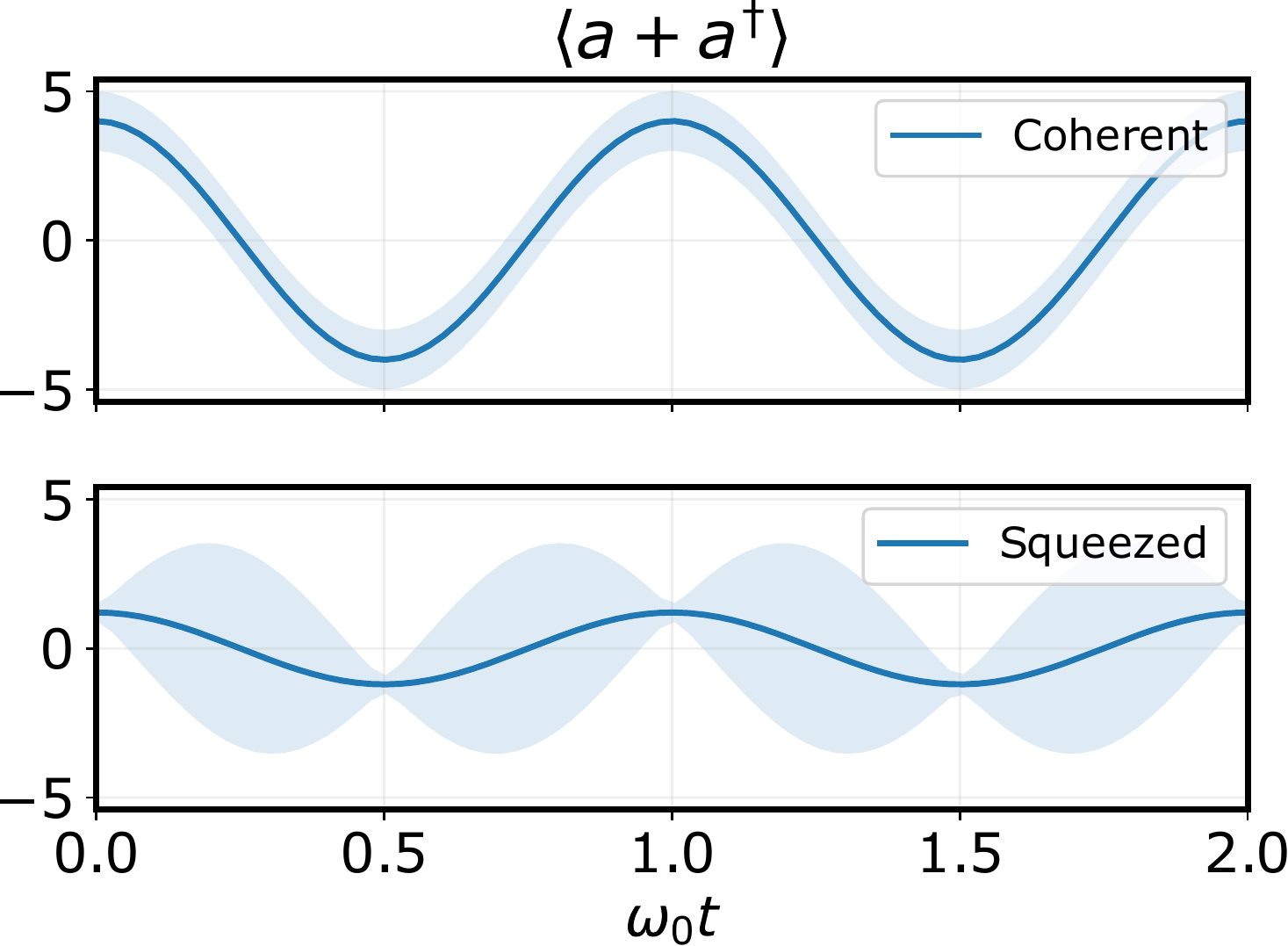}
    \caption{Time evolution of mean (solid line) and variance (shaded region) of the field quadrature for a single mode. Top: a coherent state with $ \vert \alpha \vert^{2} = 4 $. Bottom: a squeezed-coherent state with $ \vert \alpha \vert^{2} = 4 $ and $ r = 1.2, \phi = 0 $, displaying reduced noise at specific times showing that at those moments, the state is closer to an eigenstate of the electric field ($ X  $ quadrature) than a coherent state.}
\label{quadrature_examples_1}
\end{figure}

We consider two states: a coherent state with complex amplitude parameter $ \alpha  = 2 $ and a squeezed-coherent state with amplitude $ \alpha =  2 $ and squeezing parameter and phase given by $ r = 1.2, \phi = 0 $. The result of this simulation is shown in Figure \ref{quadrature_examples_1}.

\subsection{Damped harmonic oscillator} \label{sec:damped_dynamics_code}

\begin{figure}[!ht]
    \centering
    \includegraphics[width=0.8\linewidth]{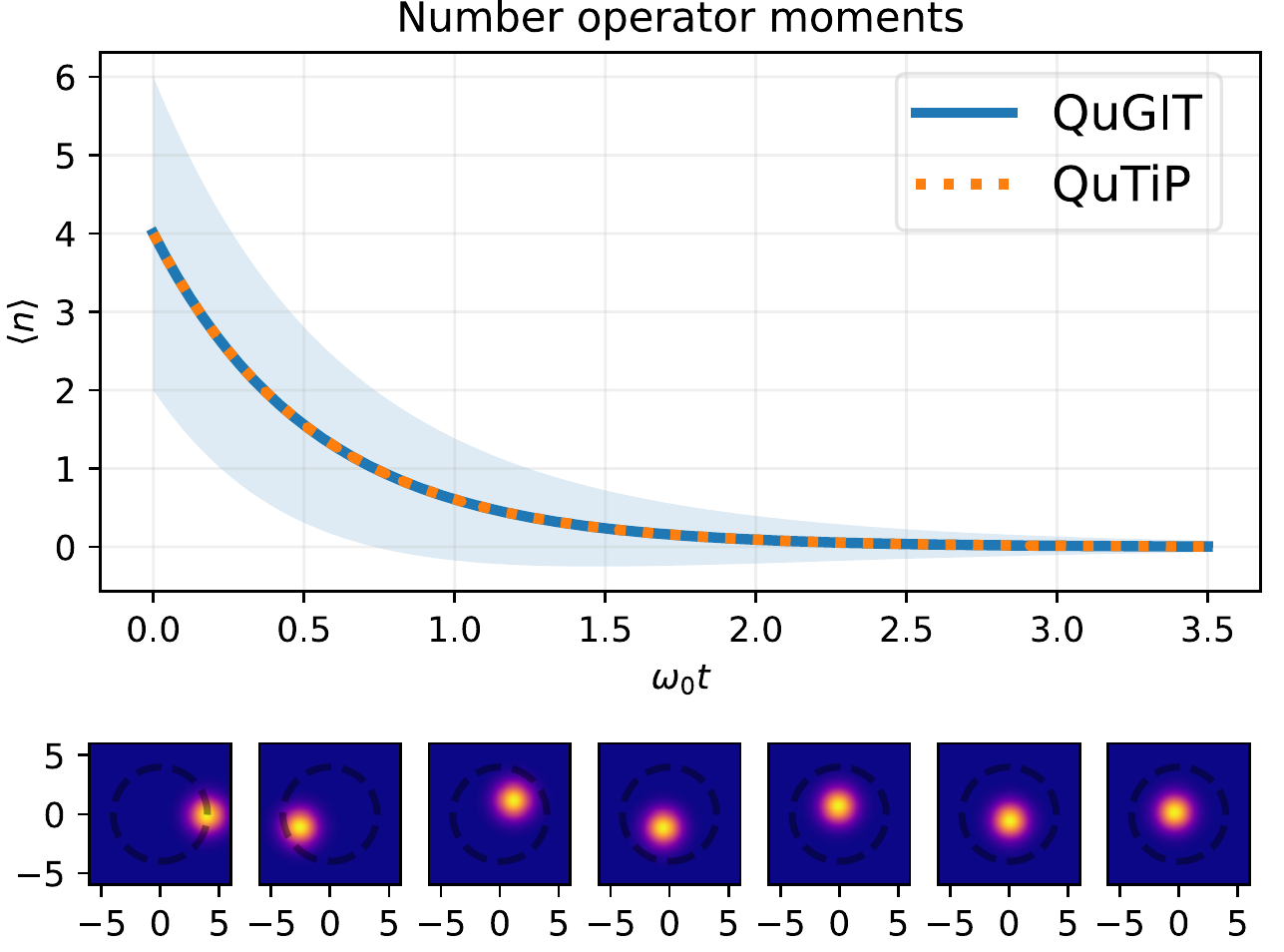}
    \caption{Damped time evolution of an initial coherent state, $\ket{\alpha=2}$. Top: occupation number calculated through QuGIT (solid blue line) compared to QuTiP (orange dots), together with the variance of the mode number operator (shaded light blue region). Bottom: Wigner functions evaluated using QuGIT; time flows from left to right. Parameters used in the simulation are $\omega_0=2\pi$ and $\gamma=2\pi\times0.3$. For QuTiP, the Hilbert space dimension truncation was $N=30$.} 
    \label{fig:damped_dynamics}
\end{figure}

We can extend the above example to include the effects of damping in the field. We consider the time evolution of a damped quantum harmonic oscillator governed by the Lindblad equation,
\begin{equation}
    \dot{\rho} = -\frac{i}{\hbar}\comm{\hbar \omega \a\dager\a}{\rho} + \gamma\left( \a\rho\a\dager - \half\a\dager\a\rho - \half \rho \a\dager\a \right) \, ,
    \label{damped_lindblad}
\end{equation}
\noindent where $\omega$ is the frequency of the mode and $\gamma$ is the damping constant. The commutator in the master equation dictates unitary dynamics, while the second term governs the interaction with the environment.  The open quantum dynamics is modelled through amplitude damping on the number of excitations of the harmonic oscillator. 
The associated Langevin and Lyapunov equations entailed by Equation \eqref{damped_lindblad} are characterized by a vanishing driving vector and the following drift and diffusion matrices,
\begin{align}
    A = \begin{bmatrix} -\gamma/2 & +\omega \\
                        -\omega   & -\gamma/2 
        \end{bmatrix} \ ,  \ \ 
    D = \begin{bmatrix} \gamma &    0   \\
                           0   & \gamma 
        \end{bmatrix}        \, .
\end{align} 
We consider the system initially in a coherent state $\ket{\alpha}$ with $ \alpha = 2 $. 
A visualization of the dynamics can be seen in Figure \ref{fig:damped_dynamics}.
On the lower plots, we show snapshots of the Wigner function for the state at different moments before a full oscillation is complete. The dashed circles represent the expected mean trajectories produced by the unitary dynamics, for comparison. One can observe that mean position of the Gaussian distribution undergoes a circular damped motion, eventually settling at the origin corresponding to the stationary vacuum state.

This damped motion of the distribution can also be visualized on the top plot, through the mean mode occupation number and its variance, both monotonically decreasing. 
For comparison we present the same simulation performed using QuTiP, dotted orange line. 
As a final check, one can also compute the quantum fidelity between the steady state of the system and the vacuum state and verify it to be 100\%.

\begin{figure}[ht!]
\centering
\includegraphics[width=0.7\textwidth]{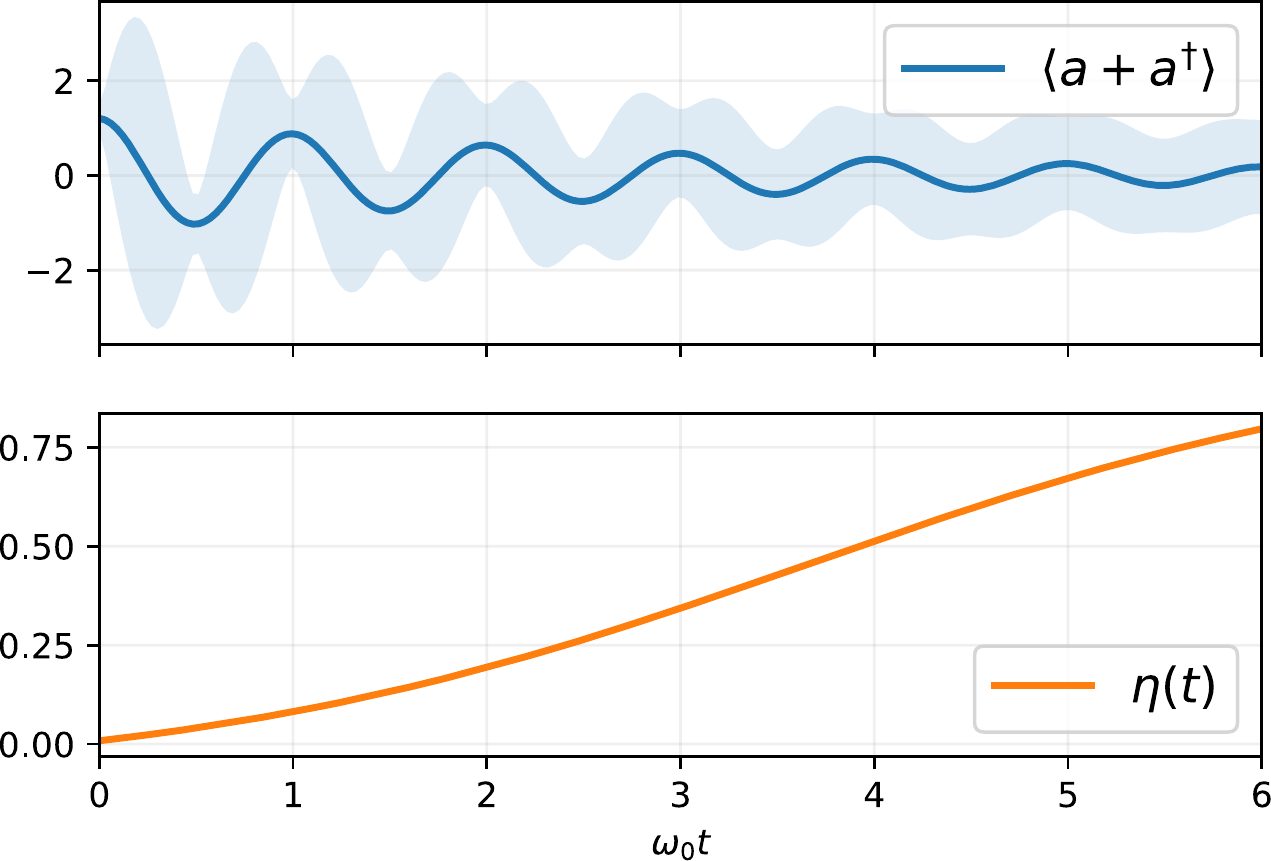}
\caption{Damped time evolution of a squeezed state visualized in terms of time-dependent field quadrature $ \langle X(t) \rangle = \langle a^{\dagger} + a \rangle $ (Top) and squeezing degree $ \eta(t) $ (Bottom). Initial squeezed state is a squeezed-coherent state with $ \alpha = 2 $ and $ r = 1.2 $. Parameters for the dynamics are: $ \omega_{0} = 2\pi $ and $ \gamma = 0.1 \times \omega_{0} $.}
\label{wigner_examples_4}
\end{figure}

We conclude this section on open quantum dynamics with an example of special interest to quantum optics, namely squeezed states. These states are sensitive to damping and photon-loss decoherence \cite{Lvovsky2016}, as we now verify using QuGIT. We simulate an analogous time evolution to the previous example, now with an initial squeezed-coherent state  $\ket{ \alpha = 2, \, r = 1.2 }$ subject to a damping constant $  \gamma = 2\pi \times  0.1 $.
On the top plot of Figure \ref{wigner_examples_4} we observe a dynamics that initially resembles the unitary evolution of the field quadrature described in the previous section. However, damping quickly acts on the field causing it to approach the vacuum. Note the reduced variance associated to squeezing of the quadrature is also affected, evidenced by a \textit{smoothing} over time.
The bottom plot of Figure \ref{wigner_examples_4} further illustrates this effect by displaying the time evolution of the squeezing degree, defined as the ratio of the squeezed to anti-squeezed field quadrature variances. The squeezing degree starts near zero for a squeezed state, and gradually evolves to unity as squeezing is degraded by the amplitude-damping dynamics. We numerically verify that in the steady state, the squeezing degree approaches 1.


\subsection{Mode entanglement and displacement detection}

A number of experiments in quantum networks use the so-called sources of heralded single photon entanglement \cite{Monteiro2015, Humphreys2018}, which consists in producing single photon states and subsequently delocalizing these over different spatial modes using beam-splitters (BS). Detection of single photon entanglement and more generally of \textit{mode-entanglement} can be achieved through the use of displacement-based detection schemes \cite{Bruno2013}, which allow for violations of Bell inequalities \cite{Guerreiro2016} as well as quantum communication protocols \cite{Monteiro2017}. 
Displacement-based detectors are Gaussian operations and as such QuGIT can be used to explore the physics of these quantum communication experiments.

\begin{figure}[t!]
\centering
\includegraphics[width=\textwidth]{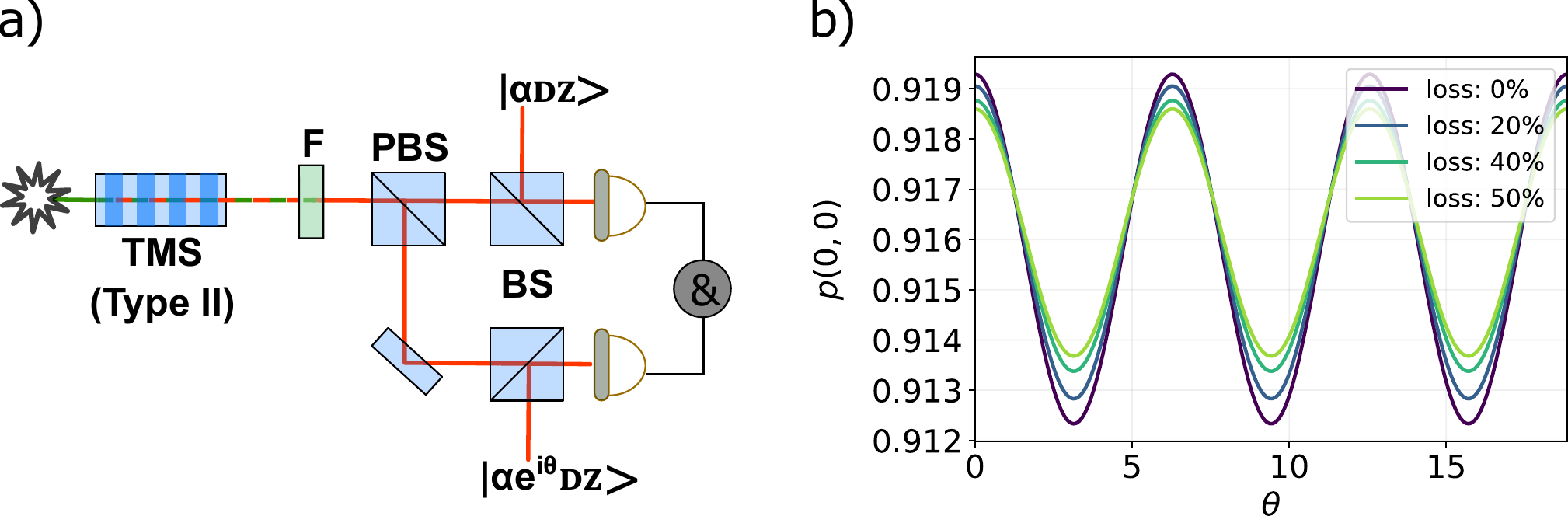}
\caption{a) Typical setup from single-photon entanglement experiments. b) Oscillations in photon-counting statistics after a displacement-based detection scheme, calculated using QuGIT. Losses are modelled through a BS interaction with ancillary vacuum modes; legend indicates transmission of the setup.}
\label{single_photon_entanglement}
\end{figure}

Figure \ref{single_photon_entanglement}a) shows a typical setup for a mode-entanglement experiment. A coherent source pumps a second-order nonlinear crystal (which can be placed inside a cavity as in the case of an optical parametric oscillator), producing Type II downconverted photon pairs. The statistics of the generated state is that of a two-mode squeezed state (TMS). The pump laser is filtered (F), and the modes in the TMS are spatially split using a polarizing beam-splitter (PBS). Each mode is subsequently input into unbalanced beam-splitters and mixed with local oscillators with a relative phase reference of $ e^{i\theta} $. The joint probability of photon counts is detected using photodiodes and coincidence logic.
In this setup, the joint photon number probability after displacement-based detection can undergo oscillations as a function of the relative phase $ \theta $. In particular, the probability of detecting zero photons at each of the photodiodes is given by,
\begin{eqnarray}
p(0,0) = \mathrm{Tr} \left( D_{a}(\alpha) D_{b}(\alpha e^{i\theta}) \vert \Psi_{TMS} \rangle \langle \Psi_{TMS} \vert D^{\dagger}_{a}(\alpha) D^{\dagger}_{b}(\alpha e^{i\theta}) \vert 0 \rangle \langle 0 \vert \right)
\end{eqnarray}
where $\vert \Psi_{TMS} \rangle$ is the TMS state and $ D_{a}, D_{b} $ are displacement operators in modes $ a $ and $ b $, respectively. Figure \ref{single_photon_entanglement}b) displays a numerical plot of $ p(0,0) $ calculated using QuGIT, taking into account losses in the interferometer through BS interactions with ancilla modes. This simple example demonstrates the power of the toolbox in simulating quantum interference experiments performed using Gaussian operations plus photon-counting, such as required in boson sampling \cite{Zhong2020} and photonic quantum computing \cite{bartolucci2021}.

\subsection{Conditional dynamics}

QuGIT is capable of solving general stochastic conditional dynamics. To demonstrate this capability, we consider the example of an Optical Parametric Oscillator (OPO) generating single mode squeezing via the Hamiltonian outlined in Section 6.1 of Ref \cite{Genoni2016}, given by
\begin{eqnarray}
\hat{H} = \frac{\chi}{2} \left(  \hat{x} \hat{p} + \hat{p} \hat{x} \right) \, ,
\end{eqnarray}
\noindent where $\chi$ is the squeezing rate.

The OPO mode can interact with an environment, modelled as a thermal bath with density matrix $V_{B} = (2 n_{th} + 1) \mathds{1}$. For simplicity, we will consider $ n_{th} = 0 $. Interaction between the bath and the OPO mode is modelled by the interaction matrix $ C = \sqrt{\gamma} \mathds{1} $, where gamma is a damping constant. The associated drift and diffusion matrices are respectively given by \cite{Genoni2016},
\begin{align}
    A = \begin{bmatrix} -\chi -\gamma/2 & 0 \\
                        0   & \chi -\gamma/2 
        \end{bmatrix} \ ,  \ \ 
    D = \gamma  (2 n_{th} + 1) \mathds{1}    \, .
\end{align} 

This defines a stable unconditional dynamics whenever $ \chi < \gamma/2$, and one can show that the steady state squeezing degree (defined as the ratio of squeezed to anti-squeezed quadratures) reads \cite{Genoni2016}
\begin{eqnarray}
\eta_{\mathrm{uncond}} = \frac{1 - 2\chi/\gamma}{1 + 2\chi/\gamma} \, .
\end{eqnarray}

We now consider the OPO mode is continuously monitored via projections onto the covariance matrix $V_{m} = \mathrm{diag}(s,1/s)$. We take the case of homodyne measurement of the $\hat{x}$ quadrature ($s \xrightarrow{} \infty$). In this case, the steady state acquires a squeezing degree given by
\begin{eqnarray}
\eta_{\mathrm{cond.}} = \left(\frac{\gamma - 2\chi}{\gamma}   \right)^{2}
\end{eqnarray}
 
\begin{figure}[!ht]
    \centering
    \includegraphics[width=0.8\linewidth]{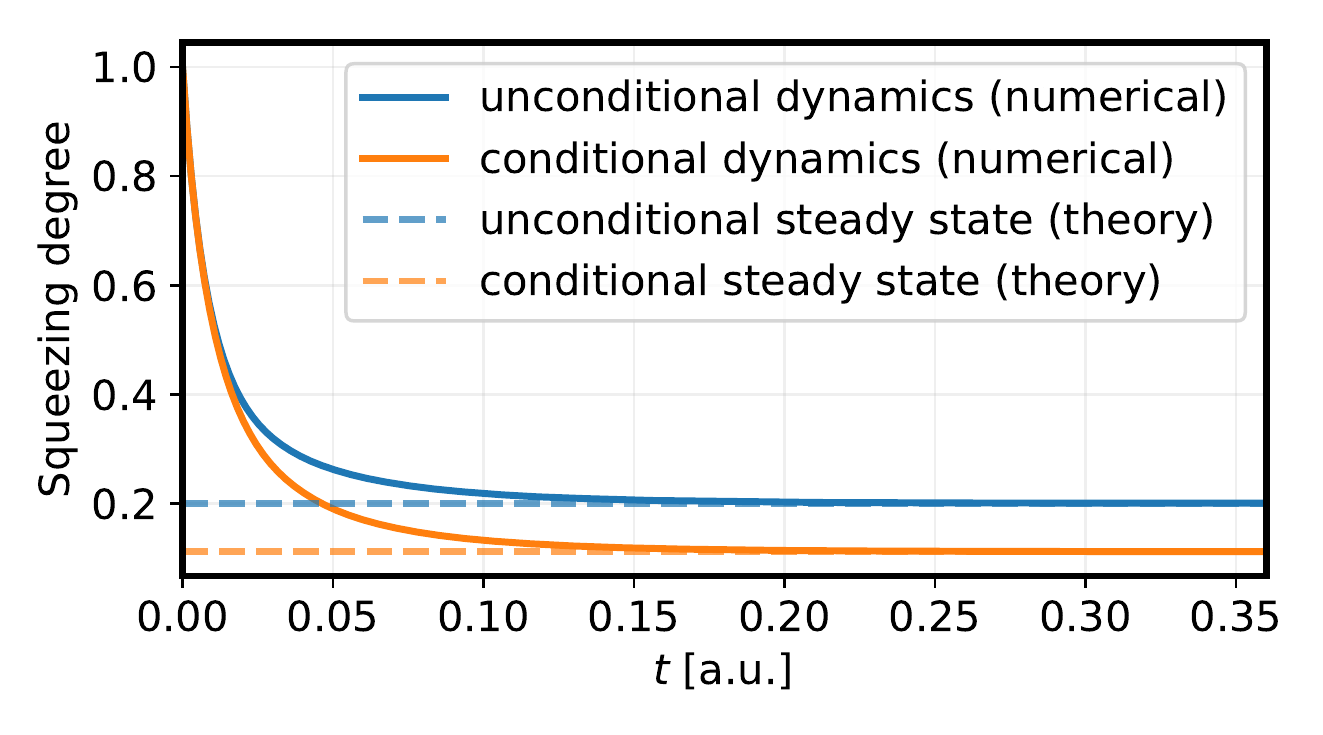}
    \caption{Unconditional and conditional dynamics of the squeezing degree of an OPO. In the conditional case, the OPO mode is subject to continuous. homodyne detection of the $\hat{x}$ quadrature. $ N_{traj} = 100$ quantum trajectories were considered for the quantum Monte-Carlo simulation.}
    \label{fig:conditional_dynamics}
\end{figure}

Figure \ref{fig:conditional_dynamics} shows the time evolution of the squeezing degree for $ \chi = \gamma/3$ for both the unconditional and conditional dynamics, together with their corresponding values in the steady state predicted by the theory. Observe that continuously monitoring the system enhances the steady state squeezing of the OPO mode.

This example illustrates the use of QuGIT to the study of conditional stochastic quantum evolution. We close this example highlighting that the above-described stochastic evolution tools can be further generalized to include feedback and optimal control according to the needs of the studied system, as for example in \cite{Magrini2020}.

\subsection{Random Gaussian circuits} \label{sec:random_circuits}



We now turn to the problem of random quantum circuits, for which the dynamics of entanglement growth has recently gained increased attention \cite{Nahum2017, Fisher}. As we next demonstrate, QuGIT can efficiently generate and simulate random unitary circuits and quantify entanglement growth under increasing numbers of gates and modes.

We generate circuits by randomly selecting a gate for each mode/pair of modes of a $N$-mode Gaussian state for $T$ turns, winding up with a quantum circuit made of $NT$ elementary gates. Elementary gates are taken from a list containing the identity, rotation, displacement, squeezing, two-mode squeezing, and beam splitter operations. When necessary, the parameter associated to each gate (such as the rotation angle for the rotation operator) are uniformly chosen from a pre-established range. 
The random circuit is then applied to a tensor product of $N$ vacuum states, typically producing a highly entangled random state.

We think of the modes as organized in a discrete 1D lattice with each mode located at integer values.
Define the spatial-dependent von Neumann entropy entropy $S(x)$ as the entropy of a bipartition of the system at mode $x$. $S(x) $ is used as a spatial-sensitive measure of entanglement, and the dynmanics of $ S(x) $ under varying number of gates quantifies entanglement growth.
Given large numbers of modes and elementary gates, QuGIT has a solid advantage in simulating examples of this type with respect to QuTIP, where trucation of Hilbert space would severely limit the simulation.

\begin{figure}[!ht]
    \centering
    \includegraphics[width=0.7\linewidth]{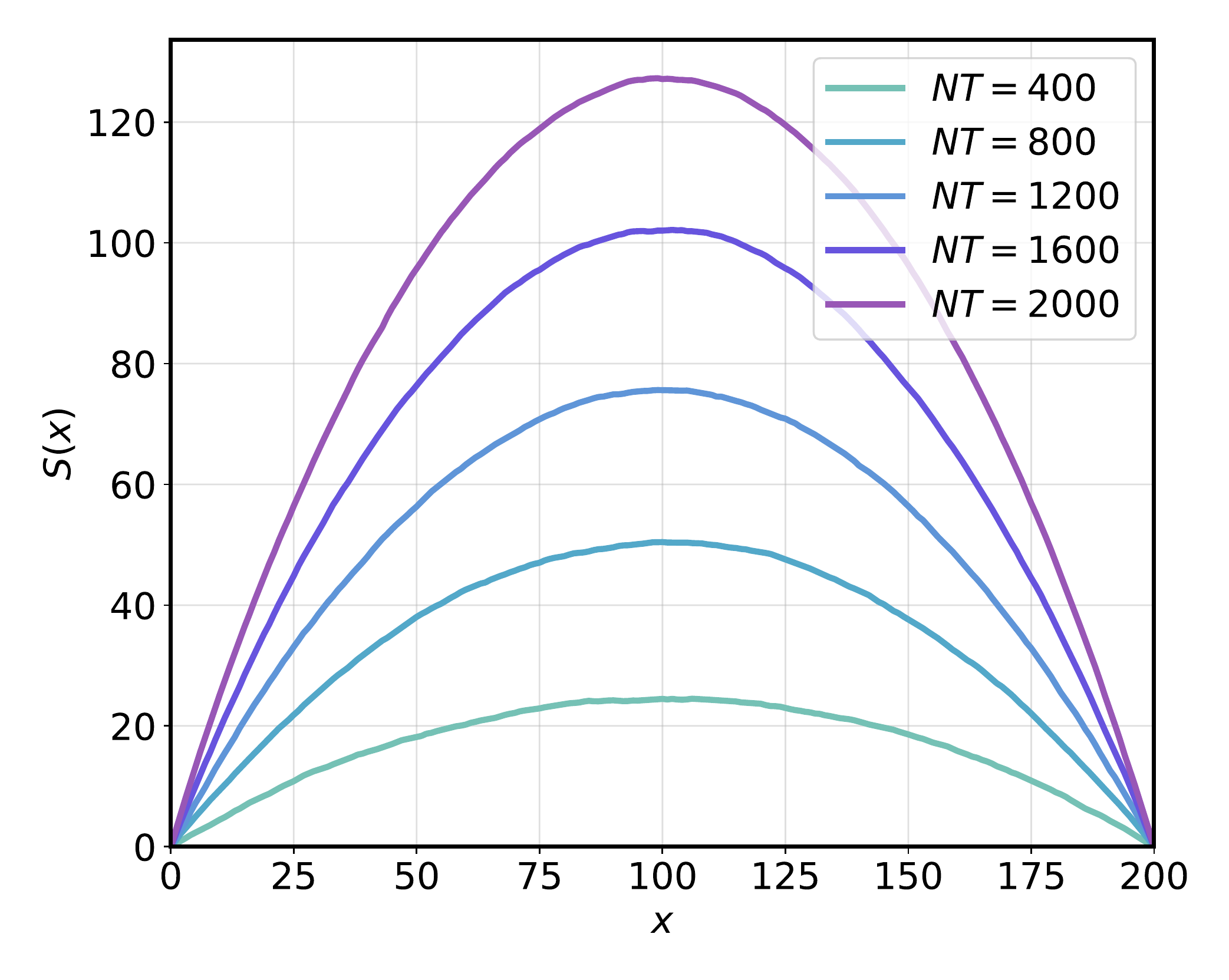}
    \caption{The spatial von Neumann entropy $S(x)$ for random Gaussian states comprised of $N = 200$ modes, obtained by applying random unitary quantum circuits to an initial vacuum state. Each circuit contains $NT$ gates. For each curve shown we average over an ensemble of $100$ realizations.} 
    \label{fig:gaussian_page_curve}
\end{figure}

Figure \ref{fig:gaussian_page_curve} displays the simulation result for $N = 200$ modes with five increasing values for $T$. 
Due to the statistical nature of this example ensembles of $100$ realizations for each value of $T$ were considered, with each curve shown expressing the mean value of $ S(x) $ over all realizations. The high value of $N$ was selected to experiment with the performance of the toolbox. In order to produce the plot in Figure \ref{fig:gaussian_page_curve}, $500$ random circuits were simulated, with an overall number of $600,000$ gates. The von Neumann entropy had to be calculated approximately $100,000$ times. The final result was obtained in 3 hours with a simple average notebook, $\SI{1.8}{GHz}$ dual-core $\SI{8}{GB}$ laptop. The outcome is consistent with simulations obtained with qubits and quantum circuits composed of Clifford gates \cite{Nahum2017} providing a continuous-variable analogue for known results of entanglement growth under random unitaries acting on finite-dimensional Hilbert spaces \cite{Page1993}.

\section{Performance } \label{sec:performance}

We now turn our attention to the performance of QuGIT. In general, the number of modes for the system in question heavily determines computation time. This can be readily observed when considering the time evolution of systems with a large numbers of interacting modes.
Consider the problem of calculating the open quantum dynamics of a system of $N$ modes in which all modes interact with independent thermal baths and with one another such that all mode-quadratures are coupled, i.e. the matrix $H(t)$ in Equation \eqref{eq:gaussian_Hamiltonian} has no vanishing elements. In this case the unitary dynamics is governed by the Hamiltonian
\begin{equation}
\hat{H} = \hbar\sum_{j=1}^N \omega_j \hat{b}_j\dager\hat{b}_j + \sum_{\substack{j=1\\k\neq j}}^N \alpha_{jk} \hat{x}_j\hat{x}_k + \beta_{jk} \hat{p}_j\hat{p}_k + \gamma_{jk} \hat{x}_j\hat{p}_k + \delta_{jk} \hat{p}_j\hat{x}_k \, ,
\end{equation}
\noindent while interaction with the environment is modelled as an independent thermal bath at finite temperatures for each particle, inducing quantum Brownian motion on each mode. Here, we refer to Ref \cite{Giovannetti2001} for the equations of motion dictating this Gaussian-preserving open system dynamics.

\begin{figure}[!ht]
    \centering
    \includegraphics[width=0.6\linewidth]{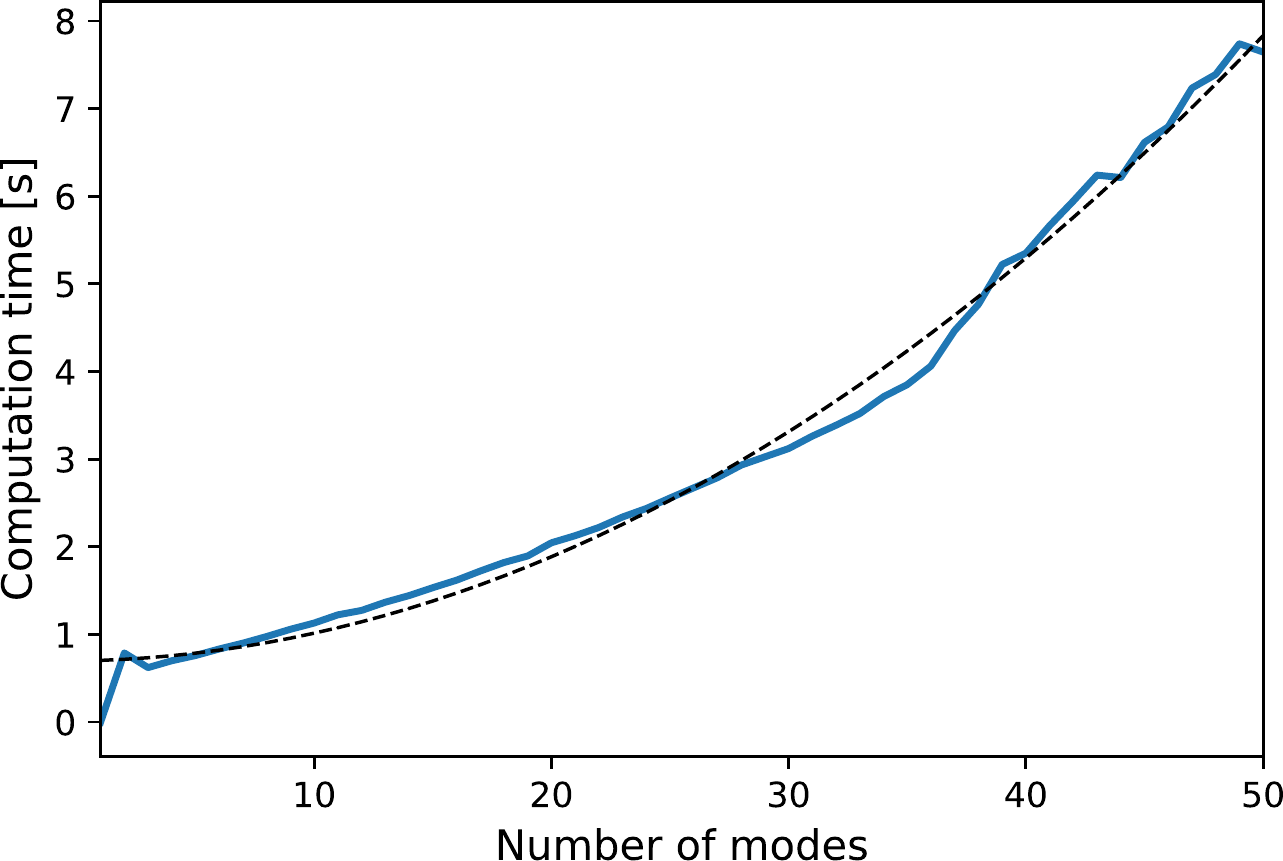}
    \caption{Blue: Computation time for a generic unconditional dynamics, averaged over $50$ realizations, as a function of the number of modes. Each simulation consisted of a time evolution of $10,000$ time steps spanning 5 complete cycles of the harmonic oscillators. Black dashed line: quadratic fit. Parameters used were $\omega_j=2\pi\times\SI{305}{kHz}$, and $\alpha_{jk}=\beta_{jk}=\gamma_{jk}=\delta_{jk}\sim\omega_j/3$.}
    \label{fig:performance_unconditional_dynamics_simple}
\end{figure}

Figure \ref{fig:performance_unconditional_dynamics_simple} shows the computation time needed to calculate the time evolution of the system using QuGIT as a function of the number of interacting modes, averaged over $50$ realizations. Note the polynomial scaling with the number of modes. 
It is instructive to compare computation times for QuTiP versus QuGIT.
While QuGIT could exactly simulate a $50$-mode Gaussian dynamics in $\SI{8}{s}$, QuTiP takes $\SI{25}{s}$ to simulate a similar $4$-mode Gaussian dynamics with a Hilbert space of dimension of $5$.
It is expected that in the ideal case the toolbox performance scales as $N^2$, following the growth in size of the covariance matrix for Gaussian states. Figure \ref{fig:performance_unconditional_dynamics_simple} includes a quadratic fit to the computation time displaying good agreement with the data for $N\leq 50$. 

\begin{figure}[!ht]
    \centering
    \includegraphics[width=0.7\linewidth]{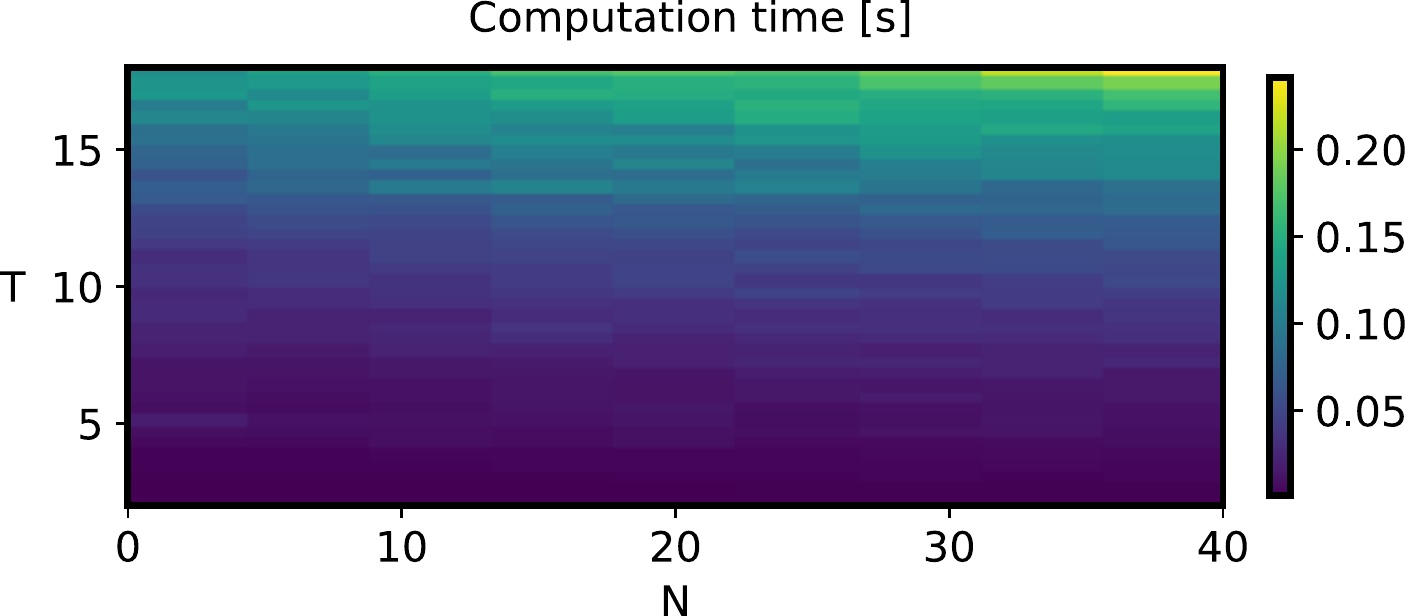}
    \caption{Computation time for random Gaussian circuits as a function of the number of modes $N$ and number of gates per mode $T$.}
    \label{fig:performance_random_circuits}
\end{figure}

As a second example of performance, we revisit Section \ref{sec:random_circuits} and study the computation time of QuGIT for random circuits. 
The total computation time including state initialization, application of the random quantum circuit and calculation of the spatial-dependent von Neumann entropy is shown in Figure \ref{fig:performance_random_circuits} as a function of number of modes and gates.
Observe that the number of gates per mode has a more significant impact on the performance over the number of modes.
For the most challenging case of a total of $NT=720$ gates applied to $N=40$ modes the simulation completes in less than a third of a second on the average laptop computer. 
All performance tests were carried out on a $\SI{2.80}{GHz}$ quad-core, $\SI{16}{GB}$ laptop running Windows.

\section{Conclusion} \label{sec:conclusions}

In this work, we report an open-source numerical Python toolbox to simulate Gaussian quantum states and operations. By directly using the symplectic representation of Gaussian states, the toolbox can exactly simulate multi-mode systems without the need for truncated Hilbert spaces or other approximations. The resources needed to store and manipulate quantum states in QuGIT is greatly reduced prompting the use of the toolbox to simulate systems with many constituents.

Various numerical examples were carried out to exhibit the toolbox versatility and robustness for a variety of Gaussian quantum systems and dynamics. The performance of the toolbox was considered. In conclusion, while the quantum information community benefits from excellent packages for simulating quantum systems such as QuTiP, we hope QuGIT will add to that list, providing complementary solutions when it comes to the simulation of Gaussian continuous variable systems.



\section*{Citation guideline}

If you make use of QuGIT in your research please add a citation to this paper and acknowledge using:
\begin{center}
\textit{This work makes use of the QuGIT toolbox.}
\end{center}

\section*{Acknowledgements}
We thank Pedro Paraguass\'{u}, Bruno Suassuna and Igor Califrer for helpful discussions and testing of the toolbox. I.B. thanks Dr. Dan Marchesin for the numerous discussions on efficient programming. This work was financed in part by Coordenaç\~ao de Aperfei\c{c}oamento de Pessoal de N\'ivel Superior - Brasil (CAPES) - Finance Code 001, by Conselho Nacional de Desenvolvimento Cient\'ifico e Tecnol\'ogico (CNPq). I. B. thanks the support received by the FAPERJ Scholarship No. E-26/200.270/2020 and CNPq Scholarship No. 140279/2021-0. T.G. thanks the  support received by the FAPERJ Scholarship No. E-26/202.830/2019. D. T. thanks the support received by the CNPq Scholarship No. 132606/2020-8.

\bibliography{main.bib}

\newpage
\appendix

\section{Toolbox methods} \label{sec:gaussian_states_methods}

\begin{table}[!ht]
\small
\caption{gaussian\_state class' methods}
\label{tab:gaussian_states_methods}
\hskip-1.5cm\begin{tabular}{llc}
\hline
\hline 
Method & Description & Reference\\
\midrule
displace                & Applies a displacement       operator on a single mode Gaussian state  & \cite{Lloyd2012} \\
squeeze                 & Applies a squeezing          operator on a single mode Gaussian state  & \cite{Lloyd2012} \\
rotate/phase            & Applies a rotation           operator on a single mode Gaussian state  & \cite{Lloyd2012} \\
beam\_splitter          & Applies a beam splitter      operator on a two    mode Gaussian state  & \cite{Lloyd2012} \\
two\_mode\_squeezing    & Applies a two mode squeezing operator on a two mode Gaussian state     & \cite{Lloyd2012} \\
apply\_unitary          & Applies a generic unitary operator given its symplectic representation & \cite{Lloyd2012} \\
loss\_ancilla   & Applies a beam splitter operator between a desired mode and an ancilla  & --- \\
\midrule
tensor\_product         & Tensor product of two Gaussian states                             & \cite{alessioserafini2017} \\
partial\_trace          & Partial trace over some modes                                     & \cite{alessioserafini2017} \\
only\_modes             & Partial trace over all but some modes                             & \cite{alessioserafini2017} \\
\midrule
matrix\_element\_coherent\_basis & Calculates the density matrix elements on coherent state basis  & \cite{Dodonov1994} \\
matrix\_element\_number\_basis & Calculates the density matrix elements on number states basis  & \cite{Dodonov1994} \\
\midrule
purity                  & Purity                   & \cite{Lloyd2012} \\
symplectic\_eigenvalues & Symplectic eigenvalues of the covariance matrix & \cite{Lloyd2012} \\
von\_Neumann\_Entropy   & von Neumann entropy             & \cite{Lloyd2012} \\
mutual\_information     & Mutual information              & --- \\
squeezing\_degree       & Ratio of the variance of the squeezed and antisqueezed quadratures  & \cite{Cernotik2020}              \\
fidelity                & Quantum Fidelity between the two Gaussian states & \cite{Banchi2015}     \\
coherence & Coherence of a multipartite Gaussian state & \cite{Xu2016} \\
occupation\_number      & Occupation number for each mode of the Gaussian state               & ---  \\
number\_operator\_moments & Calculates means vector and covariance matrix of number operator & \cite{Vallone2019}\\
number\_statistics      & Calculates the number distribution of the Gaussian state  & \cite{Dodonov1994} \\
\midrule
wigner                  & Wigner function over a 2D grid for a single mode Gaussian state  & \cite{Lloyd2012} \\
q\_function              & Hussimi Q-function over a 2D grid for a single mode Gaussian state  & \cite{Dodonov1994} \\
\midrule
logarithmic\_negativity & Logarithmic negativity for a bipartition of a Gaussian state  & \cite{Lloyd2012} \\
\midrule
measurement\_general & Conditional state after a partial Gaussian measurement& \cite{ZhangPhDThesis} \\
measurement\_homodyne & Conditional state after  a partial homodyne measurement & \cite{ZhangPhDThesis} \\
measurement\_general & Conditional state after a partial heterodyne measurement& \cite{ZhangPhDThesis} \\
\midrule
print & Prints the Gaussian state on the console log & --- \\
copy & Creates an identical copy & --- \\
\hline\hline
\end{tabular}
\end{table}


\begin{table}[!ht]
\small
\caption{Methods of the gaussian\_dynamics class}
\label{tab:methods_gaussian_dynamics}
\hskip-1.5cm\begin{tabular}{ll}
\hline
\hline 
Method & Description\\
\midrule
unconditional\_dynamics  & Calculates the time evolution of an initial state following an unconditional dynamics\\
conditional\_dynamics    & Calculates the time evolution of an initial state following a conditional dynamics \\
steady\_state            & Calculates the steady state of an unconditional dynamics \\
semi\_classical          & Calculates the semi-classical time evolution of the mean quadratures, Monte Carlo method\\
\hline\hline
\end{tabular}
\end{table}

\newpage

\section{QuGIT example codes} \label{appendix:example_codes}

We now present the codes used on the examples of this work to illustrate the capabilities of the toolbox. For simplicity, the lines of code associated with plotting have been omitted.

\subsection{Unitary field quadrature dynamics} \label{appendix:unitary_field_quadrature_dynamics_code}

\begin{python}
import numpy as np
import quantum_gaussian_toolbox as qgt

##### Parameters for the dynamics
omega    = 2*np.pi                     # Natural frequency [Hz]
t = np.linspace(0, 2/omega, int(200))  # Timestamps for simulation

A = np.array([[    0   ,  +omega ], 
              [ -omega ,     0   ]])   # Drift matrix
D = np.diag([0, 0])                    # Diffusion matrix
N = np.zeros((2,1))                    # Driving vector

##### Simulating coherent state time evolution
initial_0 = qgt.coherent(alpha=2)      # Initial coherent state
simulation_0 = qgt.gaussian_dynamics(A, D, N, initial_0)
states_0 = simulation_0.unconditional_dynamics(t) # Simulate

##### Simulating coherent squeezed state time evolution
initial_1 = initial_0.copy()          # Copy coherent state
initial_1.squeeze(r=1.2)              # Apply squeezing operator

simulation_1 = qgt.gaussian_dynamics(A, D, N, initial_1)
states_1 = simulation_1.unconditional_dynamics(t) # Simulate

##### Retrieve information from time evolved states
mean_x_1 = np.zeros(len(t))  # List to store mean quadrature, 1st simulation
var_x_1  = np.zeros(len(t))  # List to store variance,        1st simulation

mean_x_0 = np.zeros(len(t))  # List to store mean quadrature, 2nd simulation
var_x_0  = np.zeros(len(t))  # List to store variance,        2nd simulation

for i in range(len(t)):              # Loop through time-evolved states
    mean_x_0[i] = states_0[i].R[0]
    var_x_0[i]  = states_0[i].V[0,0]
    
    mean_x_1[i] = states_1[i].R[0]
    var_x_1[i]  = states_1[i].V[0,0]
\end{python}

\newpage

\subsection{Damped harmonic oscillator code} \label{appendix:damped_dynamics_code}

\begin{center}
    Number operator moments and Wigner function dynamics for coherent state
\end{center}
\begin{python}
import numpy as np
import quantum_gaussian_toolbox as qgt

##### Parameters
omega = 2*np.pi;                    # Particle natural frequency 
gamma = 2*np.pi*0.3;                # Damping constant 
t = np.linspace(0, 3.5*2*np.pi/omega, int(200))  # Timestamps for simulation

x = np.linspace(-6,6,200)
p = np.linspace(-6,6,200)
X, P = np.meshgrid(x, p);           # Meshgrid for phase-space

##### Matrices defining the dynamics
A = np.array([[-gamma/2,  +omega],
              [-omega  ,-gamma/2]]) # Drift matrix
D = np.diag([gamma, gamma]);        # Diffusion matrix
N = np.zeros((2,1));                # Driving vector
                                               
##### Simulation
initial = qgt.coherent(alpha=2)     # Initial state
                               
simulation = qgt.gaussian_dynamics(A, D, N, initial) # Time evolution instance
states = simulation.unconditional_dynamics(t)        # Simulate

##### Retrive information from time evolved states
n_bar = np.zeros(len(t))            # List to store occupation numbers
n_var = np.zeros(len(t))            # List to store variance of number operator
W = []                              # List to store Wigner functions

for i in range(len(t)):             # Loop through time-evolved states
    n_bar[i], n_var[i] = states[i].number_operator_moments()
    W.append(states[i].wigner(X, P))

ss = simulation.steady_state()      # Steady state of the system
F = qgt.fidelity(ss, qgt.vacuum())  # Fidelity with vacuum state
\end{python}

\newpage

\begin{center}
    Quadrature and squeezing degree dynamics for coherent-squeezed state
\end{center}
\begin{python}
import numpy as np
import quantum_gaussian_toolbox as qgt

##### Parameters
omega = 2*np.pi;                    # Particle natural frequency 
gamma = 2*np.pi*0.1;                # Damping constant 
t = np.linspace(0, 6, int(200))     # Timestamps for simulation

##### Matrices defining the dynamics
A = np.array([[-gamma/2,  +omega],
              [-omega  ,-gamma/2]]) # Drift matrix
D = np.diag([gamma, gamma]);        # Diffusion matrix
N = np.zeros((2,1));                # Driving vector
                                               
##### Simulation
initial = qgt.coherent(alpha=2)     # Initial state
initial.squeeze(r=1.2)
                               
simulation = qgt.gaussian_dynamics(A, D, N, initial) # Time evolution instance
states = simulation.unconditional_dynamics(t)        # Simulate

##### Retrive information from time evolved states
squeezing_number = np.zeros(len(t)) # List to store occupation numbers

mean_x = np.zeros(len(t))           # List to store mean quadrature
var_x = np.zeros(len(t))            # List to store quadrature variance

for i in range(len(t)):             # Loop through time-evolved states
    squeezing_number[i] = states[i].squeezing_degree()[0]
    mean_x[i] = states[i].R[0]
    var_x[i] = states[i].V[0,0]

ss = simulation.steady_state()      # Steady state of the system
print(ss.squezzing_degree()[0])  # Fidelity with vacuum state
\end{python}

\newpage

\subsection{Mode entanglement and displacement detection}

\begin{python}
import numpy as np
import quantum_gaussian_toolbox as qgt

phi = np.linspace(0, 6*np.pi, int(200)) # Relative phases
tau_list = np.array([1, 0.8, 0.6, 0.5])

for j in range(len(tau_list)):
    
    bipartite = qgt.vacuum(N=2)         # Initial state
    bipartite.two_mode_squeezing(r=0.4) # Apply two-mode squeezing operator
    
    bipartite.loss_ancilla(0, tau_list[j]) # Apply BS with ancilla mode
    
    Fidelity = np.zeros(len(phi))          # List of fidelities

    for i in range(len(phi)):
        coherent1  = qgt.coherent(alpha = 0.1)
        coherent2  = qgt.coherent(alpha = 0.1 * np.exp(1j*phi[i]))
        coherent12 = qgt.tensor_product([coherent1, coherent2])
    
        Fidelity[i] = qgt.fidelity(coherent12, bipartite)
\end{python}

\newpage

\subsection{Conditional dynamics}

\begin{python}    
##### Paramters
gamma = 2*np.pi*10                      
nbar_env = 0      
chi   = gamma/3        

A = np.block([[ -chi - gamma/2 ,      0         ],            
              [      0         ,  chi - gamma/2 ]]) # Drift matrix
        
D = gamma*(2*nbar_env+1)*np.eye(2)                  # Diffusion matrix
N = np.zeros((2,1))                                 # Driving vector

initial_state = qgt.coherent(alpha=3) # Initial state
t = np.linspace(0, 0.36, 2000)        # Timestamps for simulation

##### Unconditional dynamics
simulation = qgt.gaussian_dynamics(A, D, N, initial_state)  # Simulation instance

unconditional_states = simulation.unconditional_dynamics(t) # Simulate

unconditional_sq = np.zeros(len(t))   # Calculate time evolved squeezing degree
for i in range(N_time):
    unconditional_sq_temp = qgt.squeezing_degree(unconditional_states[i])
    unconditional_sq[i] = unconditional_sq_temp[0]

###### Conditional dynamics
C = np.diag([np.sqrt(gamma), np.sqrt(gamma)]) # System-bath interaction
rho_b =  qgt.thermal(nbar_env)                # Bath's state

conditional_states = simulation.conditional_dynamics(t, N_ensemble=100,
C_int = C, rho_bath =  rho_b, s_list=[1e-5], phi_list=[np.pi/2])

conditional_sq = np.zeros(len(t))
for i in range(N_time):
    conditinal_sq_temp = qgt.squeezing_degree(conditional_states[i])
    conditional_sq[i] = conditinal_sq_temp[0]

###### Analytical predictions
a = 1 / (1 + chi/(gamma/2))
b = 1 / (1 - chi/(gamma/2))
steady_unconditional = (a/b) * np.ones(N_time) # Steady state unconditional

##### Analytical prediction for conditional dynamics
c = (gamma - 2*chi) / gamma
d = 1/c
steady_conditional = (c/d) * np.ones(N_time) # Steady state conditional
\end{python}

\newpage

\subsection{Random Gaussian circuits}

For this example, we focus on the toolbox usage and do not show the code that chooses a random Gaussian circuit and applies it to an initial state. The circuits are composed of: displacement, single-mode squeezing, rotation, beam-splitter and two-mode squeezing operators, whose parameters are chosen at random. The choice of the circuit is done by the method \pythoninline{random_circuit} and its application on a tensor product of vacuum states is carried out by \pythoninline{apply_circuit}. The code for these methods can be found at: \newline \url{https://github.com/IgorBrandao42/Quantum-Gaussian-Information-Toolbox}.

\begin{python}
import numpy as np
import random

import gaussian_toolbox as qgt

N = 200           # Number of modes for the initial state
mean_alpha = 0.1  # Mean real parameter for the displacement operator
std_alpha  = 0.01 # Standard deviation for the displacement operator's parameter
T = [2,4,6,8,10]  # Number of gates per mode to apply to the initial state
loops = 100       # Number of iterations to find the mean entropy

s_x = []          # List to store each mean von Neumann entropy
for i in range(len(T)):      # For each gate                 
    
    s_x_mean = np.zeros(N+1) # Mean entropy for this number of gates applied
    
    for j in range(loops):   # Repeat to find the average entropy
        # Generate random Gaussian circuit
        circuit = random_circuit(N,mean_alpha,std_alpha,T[i])
        
        initial_state = qgt.vacuum(N)  # Initial state
        
        # Apply random circuit to initial state
        final_state = apply_circuit(initial_state,circuit)
        
        s_x_temp = np.zeros(N+1)   # Temporary variable
        
        for k in range(1,N+1):     # Loop through each mode to generate bipartitions
            
            modes = list(range(k)) # Get indexes to first k modes
            
            # Get bipartition up to k-th mode
            partition = qgt.only_modes(final_state,modes)
            
            s_x_temp[k] = qgt.von_Neumann_Entropy(partition) # Calculate its entropy 
    
        s_x_mean = s_x_mean+s_x_temp # Add current entropy to the sum
        
    s_x_mean = s_x_mean/loops        # Get average entropy
    s_x.append(s_x_mean)             # Append it to the list of mean entropies
\end{python}

\end{document}